\documentclass[11pt]{article}
\usepackage[latin9]{inputenc}
\usepackage{geometry}
\usepackage{color}
\usepackage{graphicx}
\usepackage{setspace}
\usepackage[unicode=true,
 bookmarks=false,
 breaklinks=false,pdfborder={0 0 1},backref=section,colorlinks=true]
 {hyperref}
\hypersetup{
 linkcolor=blue,filecolor=magenta,urlcolor=blue}

\makeatletter

\usepackage{titling}
\usepackage{rotating}
\usepackage{palatino}

\usepackage{booktabs}

\usepackage{harvard}

\usepackage[titles,subfigure]{tocloft}

\usepackage{longtable}
\usepackage{tabularx}

\usepackage{array}
\newcolumntype{P}[1]{>{\centering\arraybackslash}p{#1}}

\@ifundefined{showcaptionsetup}{}{%
 \PassOptionsToPackage{caption=false}{subfig}}
\usepackage{subfig}
\makeatother

\begin{document}

\title{Coastal Flood Risk in the Mortgage Market:\\
Storm Surge Models' Predictions vs. Flood Insurance Maps\thanks{The author thanks the National Oceanic and Atmospheric Administration
for access to the SLOSH simulation data, Drechsler, Itamar, Alexi
Savov, and Philipp Schnabl for access to the bank-level data, as well
as Matthew E. Kahn, Jesse M. Keenan for comments on early versions
of the paper. The usual disclaimers apply. The author acknowledges
support from the HEC Montreal foundation.}}

\author{Amine Ouazad\thanks{Associate professor of economics, HEC Montreal. Research professorship
in Real Estate and Urban Economics, funded by the HEC Foundation.
Email: amine.ouazad@hec.ca. }}

\date{May 2020}
\maketitle
\begin{abstract}
Prior literature has argued that flood insurance maps may not capture
the extent of flood risk. This paper performs a granular assessment
of coastal flood risk in the mortgage market by using physical simulations
of hurricane storm surge heights instead of using FEMA's flood insurance
maps. Matching neighborhood-level predicted storm surge heights with
mortgage files suggests that coastal flood risk may be large: originations
and securitizations in storm surge areas have been rising sharply
since 2012, while they remain stable when using flood insurance maps.
Every year, more than 50 billion dollars of originations occur in
storm surge areas outside of insurance floodplains. The share of agency
mortgages increases in storm surge areas, yet remains stable in the
flood insurance 100-year floodplain. Mortgages in storm surge areas
are more likely to be complex: non-fully amortizing features such
as interest-only or adjustable rates. Households may also be more
vulnerable in storm surge areas: median household income is lower,
the share of African Americans and Hispanics is substantially higher,
the share of individuals with health coverage is lower. Price-to-rent
ratios are declining in storm surge areas while they are increasing
in flood insurance areas. This paper suggests that uncovering future
financial flood risk requires scientific models that are independent
of the flood insurance mapping process.
\end{abstract}
\clearpage{}\pagebreak{}

\section{Introduction}

What is the amount of mortgage credit potentially exposed to the risk
of catastrophic hurricane storm surges? Evidence suggests that flood
insurance maps published by FEMA may not provide an accurate description
of risk.\footnote{\citeasnoun{pralle2019drawing} suggests that three quarters of houses
damaged during hurricane Harvey were outside of the 100-year floodplain;
and that half of the buildings in New York City affected by Sandy
were outside of the 100-year floodplain. \citeasnoun{kousky2018financing}
provides a discussion of the design of Flood Insurance Rate Maps.
\citeasnoun{kousky2010improving} argues that better flood maps are
required in St Louis.} New estimates of floodplain boundaries~\cite{wing2018estimates}
suggest that up to 41 million Americans live within the 100-year floodplain,
substantially above the number of Americans living within the 100-year
floodplain of FEMA's flood insurance maps. As statistics from the
Federal Reserve's Flow of Funds suggest that Americans owed about
11.2 trillion dollars of mortgage debt in 2019~\cite{goodman2020housing},
the exposure of lenders, securitizers, and households may be higher
than suggested by flood insurance maps. Flood risk may cause defaults
or prepayments among borrowers, and cause losses among lenders and
securitizers. Yet, estimating the exposure of the mortgage market
is challenging as (i)~not all communities participate in FEMA's National
Flood Insurance Program and thus flood risk may not be mapped comprehensively;
(ii)~in participating communities, the observed frequency of flooding
due to precipitation events, fluvial flooding, or hurricane storm
surges may not match the predicted frequency of FEMA's National Flood
Hazard Layer; (iii)~parts of the 100-year floodplain assume protection
by a levee; and (iv)~publicly available data may not include simple
measures of mortgage structure or performance for loans outside of
the conventional loan single-family market.

This paper provides a transparent and replicable assessment of flood
risk in the mortgage market by matching the numerical simulations
of a model of hurricane storm surges with individual mortgage files,
household demographics, house prices and rents, and lenders' balance
sheets. Such simulations predict storm surge height at a granular
level, an assessment of flood risk that is independent of communities'
willingness to participate in the flood insurance program. Such model
acknowledges the impact of levees, yet does not exclude leveed areas
from the simulation. Relying on scientific models rather than on flood
insurance maps is key, as flood insurance maps are the outcome of
a political economy process of community participation in the program
as well as of community investment in mitigation efforts. Hence scientific
models help us tease out the flood risk that is driven by climate
and weather processes separately from what is driven by social and
political interactions. Mortgage originations in areas exposed to
storm surges are 26\% higher than the total volume of originations
in FEMA's 100-year floodplains. The paper finds that areas exposed
to storm surge risk that are not in flood insurance maps' 100-year
floodplain tend to be substantially more African American and Hispanic,
exhibit price trends that are lower than the national price trend
\textendash{} while prices tend to increase faster in FEMA's 100-year
floodplain \textendash . Borrowers in storm surge areas tend to take
more interest-only and fewer fixed rate mortgages. The risk of storm
surges is taken on by larger banks, with lower return on equity, less
dependence on mortgage loans; lenders are more likely to be commercial
banks than in other areas.

Hurricane storm surge height is derived at the neighborhood-level
using the National Oceanic and Atmospheric Administration's (NOAA)
Sea, Land, and Overland Surges from Hurricanes (SLOSH) model, for
different hurricane categories and for different tide levels. The
output of this model is combined with the United States Geological
Survey's (USGS) Digital Elevation Model satellite measures of elevation,
yielding a predicted surge height above ground level. Storm surge
heights are matched with mortgage market data. Mortgage-level information
comes from two sources: public data collected by the Federal Financial
Institutions and Examinations Council (FFIEC) according to the Home
Mortgage Disclosure Act; and confidential data on mortgage structure
from the McDash data, a product of Black Knight financial. Mortgage-level
information is matched to lenders' balance sheets, lenders' liquidity
and capitalization, from the Federal Reserve of Chicago's Commercial
Bank data. As such the assessment presented in this paper presents
a novel replicable methodology that is mostly replicable using publicly
available data.

The paper's findings cast a light on the importance of measuring flood
risk using models based on weather phenomena, elevation, obstacles,
and land cover rather than by relying on flood insurance maps. An
extant literature has used storm models on a limited scale to assess
coastal flood risk~\cite{mercado1994use,melton2010hurricane,niedoroda2010analysis,genovese2015assessment}.
This paper uses the 35 basins of storm surge simulations for all Zip
code areas at the national levels to study the correlation between
the mortgage market's vulnerabilities and storm surge risk.

Recent work highlights the evolution of financial markets in response
to climate risk~\cite{hong2020climate}. By providing a unique US-wide
snapshot of current aggregate risk in the mortgage market, the paper
aims at providing ``early warning'' information to households, financial
institutions, securitizers, and other players to help in the adaptation
of the growing threat of climate risk. Indeed, climate risk is not
a static concept~\cite{kahn2014climate,kahn2016climate}. As households
and lenders learn about the risk \textendash{} including those described
in this reproducible analysis with publicly available data \textendash ,
lenders adjust their lending standards and households adapt their
borrowing behavior to the looming issue of growing defaults. Hence
this descriptive piece is not a forecast, but rather information that
helps financial markets shift the equilibrium towards a more climate
resilient mortgage portfolio.\footnote{\citeasnoun{keenan2019climate} introduces climate adaptation in asset
management (chapter 2) and in funding and financing options (chapter
4).} A future test of climate adaptation in the mortgage market is to
test the impact of such information disclosures on lenders' underwriting
standards and household demand for mortgage credit.

The paper proceeds as follows. Section~\ref{sec:Data-Set} describes
the data sets, their strengths and limitations, and the methodology
that matches them to mortgage data. Section~\ref{sec:Six-Facts}
then presents the core six facts of this paper. Section~\ref{sec:Conclusion}
concludes.

\section{Data Set\label{sec:Data-Set}}

Assessing the vulnerability of real estate assets, mortgage debt,
and lenders' financial statements to coastal flood risk requires four
sources of data. First, data on the impact of hurricane storm surge
heights or sea level rise at a fine-grained geographic level. Second,
data on individual mortgages' location, loan-to-value or equity, amortization.
Third, information on households' and individuals' vulnerability.
This includes household income, the type of dwelling (mobile home),
minority status, education, and other relevant variables that are
correlated with a household's ability to repay the mortgage. Fourth,
information on lenders' vulnerability, by matching mortgage originations
to lenders' status (bank or non-bank lender), net income, balance
sheet. We describe these information sources one by one. 

\subsection{Coastal Flood Risk}

\paragraph*{Neighborhood-Level Hurricane Storm Surge Heights}

NOAA's Sea Lake and Overland Surge from Hurricanes (SLOSH) simulates
the impact of hurricanes on storm surge heights using the transport
equations and hurricanes' measures. The initial conditions of such
equations are hurricanes' characteristics such as pressure, speed,
and track. The transport equations have been present in the literature
at least since \possessivecite{ekman1902om} seminal work, published
in English in \citeasnoun{ekman1905influence}, which were not initially
used for the forecasting of storm surge heights.\footnote{Ekman was a student of the founder of modern meteorology Vilhelm Bjerknes.}
A practical application to the modeling of storm surge dynamics is
presented later in the work of \citeasnoun{jelesnianski1970bottom}.
In this latter work, the model is tested for three storms of the Atlantic
Seaboard on Atlantic City, NJ: the September 1944 Storm, hurricane
Donna, and hurricane Carol. The availability of \emph{measured }surges
using gage records helps in comparing the predictions of the equations
with the realizations of storm surge heights. In these three key examples,
the model performs well and, when deviating from the observations,
tends to underestimate storm surge heights. \citeasnoun{jelesnianskyshaffer}
compares measures from 570 tide gage and high water mark observations
with the model's predictions for a larger set of storms. It reports
that the model's predictions are within $\pm20\%$ of observed heights.
A recent assessment of model driven forecasts of storm surge heights
is presented in \citeasnoun{kalourazi2019simulating}.

Ekman transport equations are at the core of the SLOSH model, which
has become a central tool for NOAA's National Hurricane Center forecasts
of storm surge heights~\cite{glahn2009role}. Parameters provided
by hurricane forecasters lead to finite-difference simulations of
the Ekman transport equations using storm position, the radius of
maximum winds, and the pressure difference between the central and
the peripheral pressure. \footnote{Historical measures every few hours are provided in the National Hurricane
Center's HURDAT2 data set.} 

Using model-driven predictions of storm surge heights rather than
historical observations of water gage levels is key. There is indeed
evidence that hurricane intensity has been increasing~\cite{kossin2020global},
consistent with the prediction of numerical models that a warmer world
leads to a higher intensity of hurricanes. Hence using historical
observations of storm surge heights may not be a good indication of
future storm surge risk. In this paper, we use SLOSH simulations to
provide estimates of worst-case scenarios that may materialize as
hurricanes become more intense and sea levels rise. 

SLOSH predicts storm surges in feet as the Maximum Envelope of Water
(MEOW) for a given hurricane, for instance a Hurricane with the characteristics
of Katrina in 2005. By finding the maximum of storm surge height over
possible hurricane characteristics, NOAA obtains the Maximum of MEOWs
(MOM) for a given hurricane intensity on the Saffir-Simpson scale.
MOM simulations are provided at a granular level: the MOM for the
New York basin include 30,832 individual geographic cells, each with
a storm surge prediction in feet. The cells are defined in polar coordinates.
Smaller cells are about 250m by 250m. We intersect such cells with
the finest possible level of geographic disaggregation: either the
boundaries of the US Census Bureau's census tracts in their 2010 definitions
(when working with Census demographics or public mortgage data), or
the boundaries of Zip Code Tabulation Areas 5 (ZCTA5) (when working
with confidential mortgage data, where data is available at the 5-digit
ZIP code\footnote{There are minor differences between the boundaries of ZCTA5s and the
boundaries of postal ZIPs.}).

The MOM simulations are obtained for 35 different basins,\footnote{The current version of this paper does not consider the basins of
Puerto Rico, the Virgin Islands, and Hawaii for computational reasons.} from the Penobscot Bay in Maine, down South to the Laguna Madre basin
at the border between the United States and Mexico. Such basins include
the New Orleans basin, as well as basins for Palm Beach and Biscayne
Bay, among others. Basins cover all parts of the Atlantic coastline
and coastal areas of the Gulf of Mexico. When basins overlap, we take
the maximum of the MOMs for the two basins.

\paragraph*{Sea Level Rise Forecasts}

Our second source of coastal flood risk is the series of Sea Level
Rise layers provided by NOAA. In its ``Global Sea Level Rise Scenarios
for the United States National Climate Assessment,''~\cite{parris2012global}
NOAA states that ``we have very high confidence (\textgreater{}9
in 10 chance) that global mean sea level will rise at least 0.2 meters
(8 inches) and no more than 2.0 meters (6.6 feet) by 2100.'' Using
a survey of 90 experts from 18 countries, \citeasnoun{horton2014expert}
reports that the consensus forecast is 0.7-1.2m (2.3-3.9ft) in the
unmitigated warming scenario, and 0.4-0.6m (1.3-2.0ft) in the mitigated
scenario by 2100.

Consistent with these forecasts, we use NOAA's SLR water levels ranging
from 0 to 6 feet, where such levels are relative to the local Mean
Higher High Water datum. NOAA also provides 7-10 feet SLR forecasts
for some parts of the coast, but we do not consider these in this
analysis; they would also fall outside the range of forecasts of SLR
in 2100. Both the East and the West coast SLR forecasts are considered,
from Washington to California and from Maine to Texas. SLOSH basins,
in comparison, only consider the East coast. 

Sea level rise should lead to higher storm surges~\cite{shepard2012assessing,woodruff2013coastal}.
Yet SLOSH storm surge simulations are performed at given sea level.
This paper may thus underestimate the coastal exposure of mortgages
to flood risk. In this paper we use SLOSH simulations at high tide
to compensate for the lack of SLR in the simulations.

\paragraph*{Flood Insurance Rate Maps: FEMA's Special Flood Hazard Areas}

The third source of information on coastal flood risk is FEMA's Flood
Insurance Rate Maps. In such FIRM maps, the Special Flood Hazard Areas
are strict boundaries of the 100-year floodplain, in which the annual
probability of a flood is 1\%. This paper uses the 2017 National Flood
Hazard Layer. 

Flood Insurance Rate Maps are relevant for the vulnerability of the
mortgage market to flood risk. First, households borrowing using an
agency-guaranteed mortgage are required to buy flood insurance since
the 1973 Flood Disaster Protection Act.\footnote{Section 103, (3), (B) ``GOVERNMENT-SPONSORED ENTERPRISES FOR HOUSING.-{}-The
Federal National Mortgage Association and the Federal Home Loan Mortgage
Corporation shall implement procedures reasonably designed to ensure
that, for any loan that is-{}- {[}...{]} purchased by such entity,
the building or mobile home and any personal property securing the
loan is covered for the term of the loan by flood insurance in the
amount provided in paragraph (1)(A).''} Second, 

Yet, such FIRM maps have some significant limitations. First, they
provide a strict binary boundary for the 100-year floodplain, which
may give a false sense of security for borrowers in the immediate
vicinity of the external boundary of the 100-year floodplain. In contrast,
SLOSH models provide smooth predictions of hurricane storm surges.
Measures from water gages are continuous as well. Second, some communities
do not participate in the National Flood Insurance Program, which
leads to (i)~households' inability to purchase federal flood insurance
and (ii)~the absence of flood mapping.\footnote{The list of participating communities is presented in the ``\href{https://www.fema.gov/national-flood-insurance-program-community-status-book}{National Flood Insurance Program Community Status Book}.''}
We compare flood zones below. 

\paragraph*{Comparing Flood Zones}

\begin{figure}
\caption{Two Approaches to Flood Zones: Storm Surge model and Flood Insurance
Maps\label{fig:mapping_flood_zones}}

\emph{The upper panel shows a map of the simulated Maximum of MEOWs
storm surge heights for a Category 4 hurricane at high tide. The storm
surge heights in feet are above ground level. MEOW: Maximum Envelope
Of Water. This is presented for the basin of New Orleans. The grey
area is the New Orleans-Metairie, LA Metropolitan Statistical Area.
The black boundaries are the Zip code tabulation areas (ZCTA5) used
in the mortgage analysis. The bottom panel shows the boundaries of
the 100-year floodplain in the National Flood Hazard Layer, provided
by FEMA in 2017. }

\bigskip{}

\subfloat[Storm Surge Simulation \textendash{} Category 4, High Tide \textendash{}
Surge Above Ground Level in Feet]{

\includegraphics[scale=0.2,trim={0cm 18cm 0cm 18cm},clip]{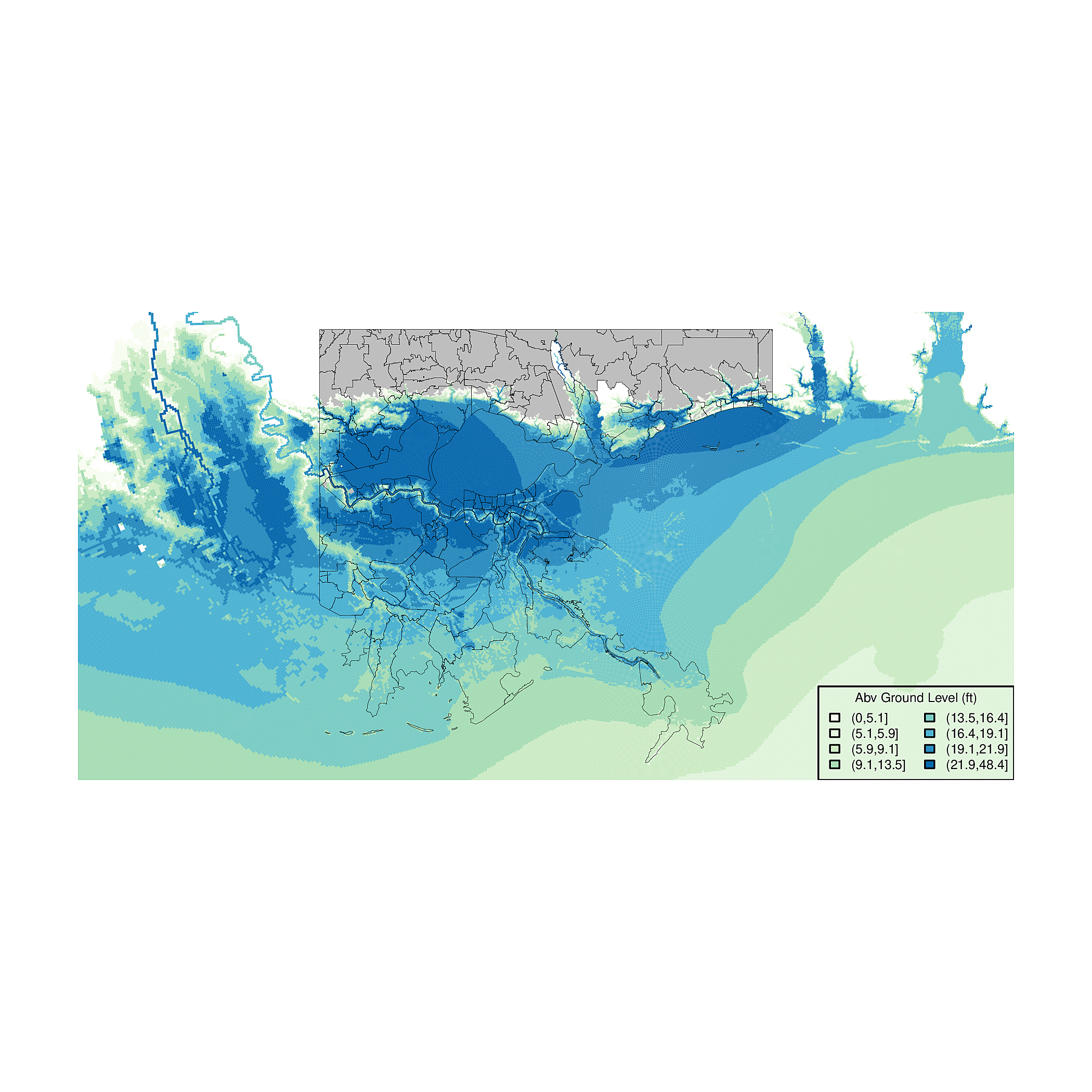}}

\subfloat[Flood Insurance 100-Year Floodplain \textendash{} Federal Emergency
Management Agency's National Flood Hazard Layer]{

\includegraphics[scale=0.2,trim={0cm 18cm 0cm 18cm},clip]{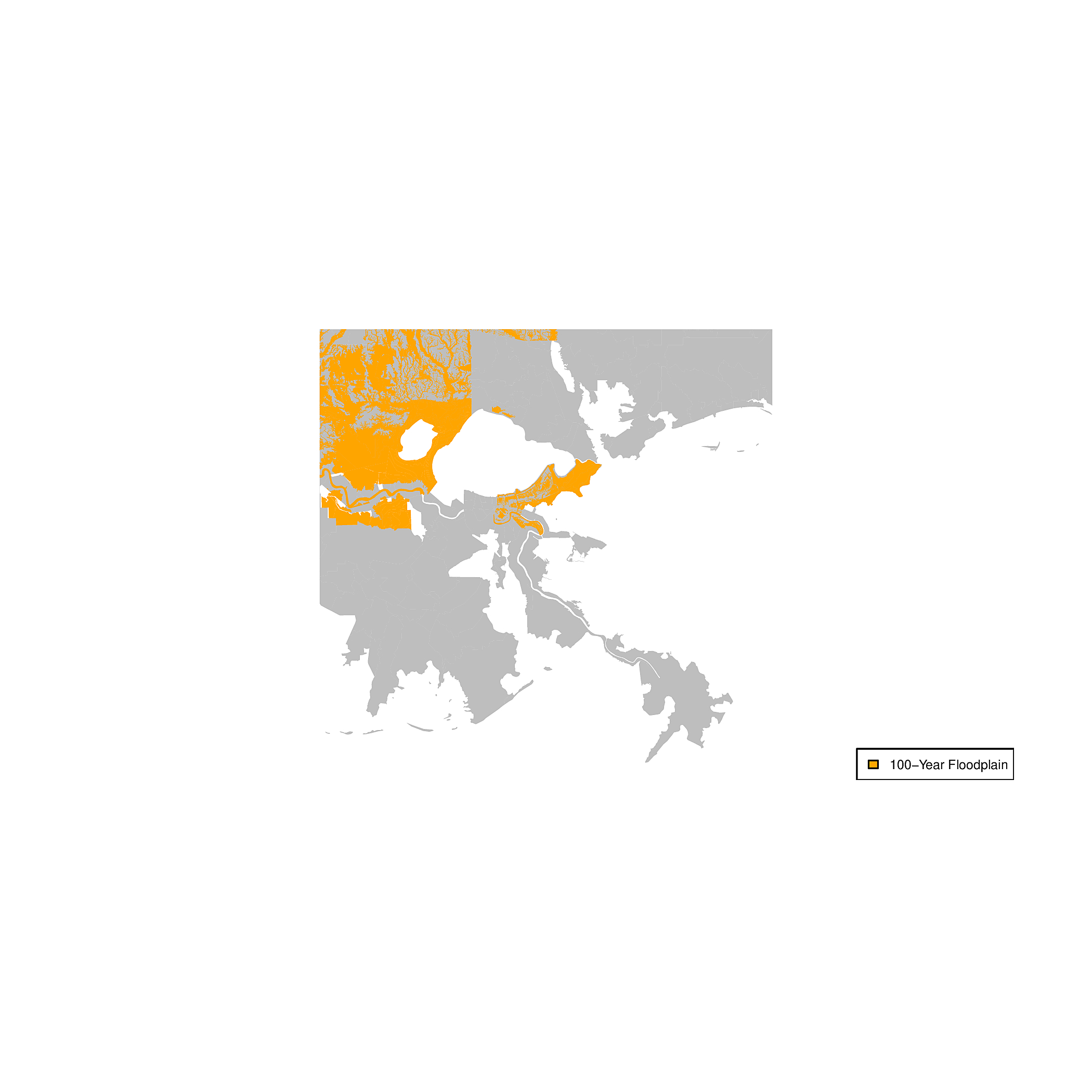}}

\bigskip{}

\emph{These figures focus on New Orleans. The paper considers the
comprehensive set of basins of storm surge simulations. Source: mapping
by the author using NOAA SLOSH simulations, FEMA NFHL, and US Census
Bureau shapefiles.}
\end{figure}

Figure~\ref{fig:mapping_flood_zones}(a) maps the storm surge areas
obtained using a Category 4 hurricane at high tide in the New Orleans
basin. Hurricane Katrina was a Category 5 floodplain, so this could
be considered a moderate scenario. Figure (b) presents the areas of
the flood insurance maps' 100-year floodplain. 

This visual representation sheds lights on the potential limits of
using the flood insurance maps to assess flood risk in the mortgage
market. First, while a Category 4 storm surge would affect all parts
of the New Orleans metropolitan area except its northern part, the
flood insurance areas cover a substantially smaller part of the metropolitan
area. This is the National Flood Hazard Layer provided by FEMA in
2017. The discrepancies are substantial: a stretch from New Orleans
to Laplace is unmapped, while the City of Kenner is assumed to be
protected by a levee and is thus outside of the 100-year floodplain.
The southern part of the state of Louisiana is mostly unmapped.\footnote{Levee reliability is the focus of a complex and extent literature
with both engineering, cost-benefit, and risk preference considerations,
see for instance \citeasnoun{tobin1995levee}, \citeasnoun{wolff2008reliability},
\citeasnoun{rogers2015interaction} .} Second, storm surge simulation models provide a continuous and granular
visualization of storm surge risk, with surges ranging from no surge
(in the northern counties of the New Orleans MSA) to more than 22
feet (in the city of New Orleans, on the Eastern side of Lake Pontchartrain). 

\begin{figure}
\caption{Sea Level Rise Scenario: 3 Feet\label{fig:Sea-Level-Rise}}
\bigskip{}

\emph{This panel present the areas flooded in the case of a 3-feet
sea level rise for the New Orleans-Metairie, LA Metropolitan Statistical
Area. }

\bigskip{}

\subfloat[3 Feet of Sea Level Rise, NOAA's Simulations, Louisiana]{

\includegraphics[scale=0.25,trim={0cm 18cm 0cm 18cm},clip]{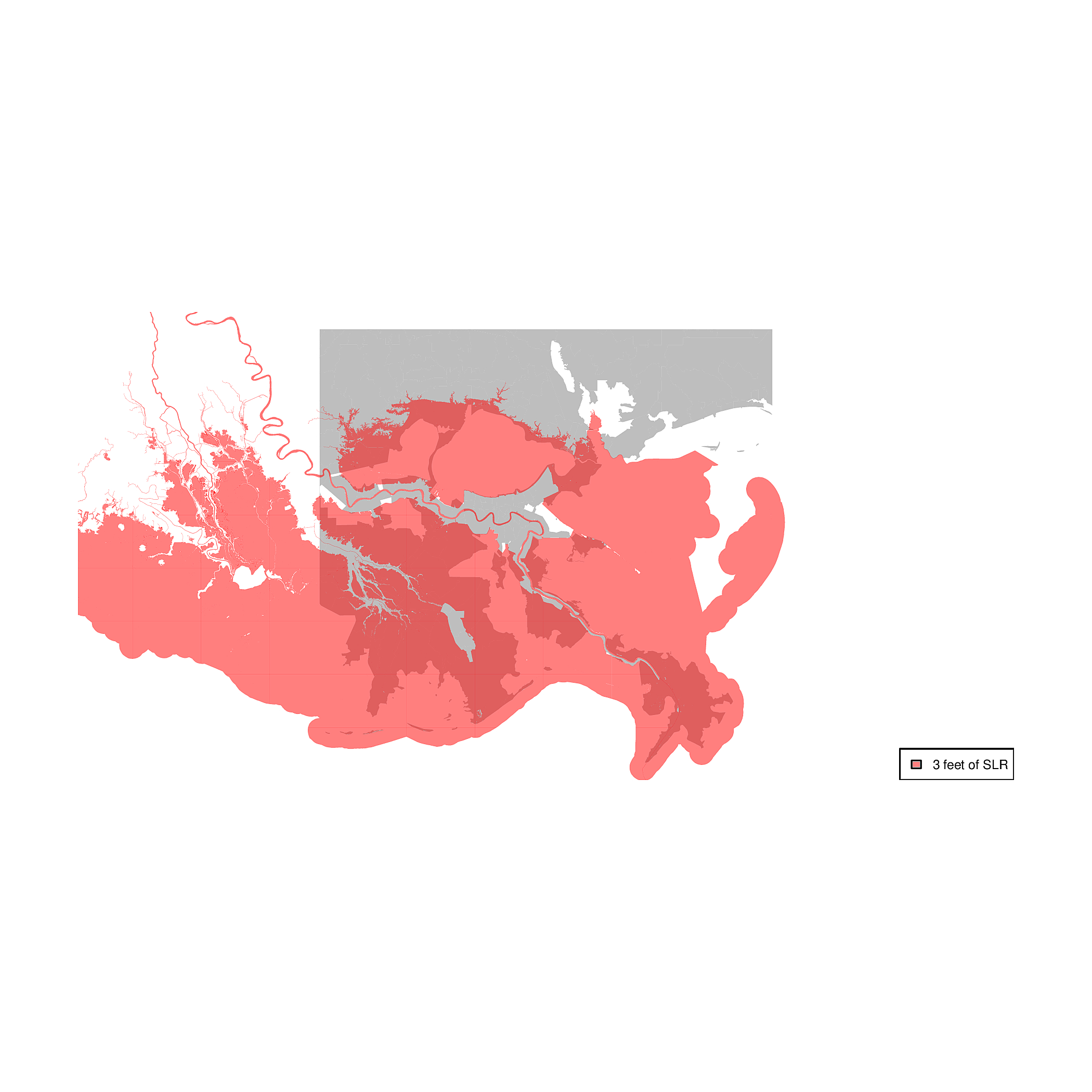}}

\emph{Source: Mapping by the author using US Census Bureau and NOAA
shapefiles.}
\end{figure}

Figure~\ref{fig:Sea-Level-Rise} on the next page presents the 3
feet sea level rise scenario; in New Orleans the 6 feet and 3 feet
scenarios are virtually identical at this scale. Such layers also
suggest that the downtown parts of the metropolitan area will not
be affected by the slow-moving sea level rise, even as they may be
affected by more frequent and higher storm surges. 6 feet of sea level
rise is on the upper bound of the likely scenarios, i.e. in the unmitigated
global warming scenarios in 2100. 

Overall, evidence suggests that NOAA's storm surge models provide
a more conservative forecast of coastal flood risk.

\paragraph*{Limits of the Measures}

This paper does not present fluvial flooding measures, which a new
version of this paper will present at a granular level for the conterminous
United States. 

\subsection{Mortgage data\label{subsec:Mortgage-data}}

This paper matches the predictions of storm surge models with instrument-level
financial and economic data.

\paragraph*{Home Mortgage Disclosure Act data}

Mortgage-level information on mortgage applications, originations
and securitizations is provided by the Federal Financial Institutions
and Examination Council and by the Consumer Financial Protection Bureau
for the period of analysis (2012-2018). Data is collected according
to the 1975 Home Mortgage Disclosure Act (codified in 12 USC Banks
and Banking). In the so-called Loan Application Register (LAR), for
each application, the data include the loan amount, the unique lender
identifier (Respondent ID), the applicant income, race, gender, ethnicity,
the census tract of the house\footnote{The file uses the Census' 2000 Tract boundaries until 2012, then adopts
the 2010 Tract boundaries. We adjust the intersections accordingly.}, the regulating agency, the loan type (conventional, FHA, VA-guaranteed,
Farm Service Agency or Rural Housing Service), the property type (1-to-4
family, manufactured housing, multifamily), the loan purpose (home
purchase, home improvement, refinancing), the owner-occupancy status.
The outcome of the application is recorded (origination, denial, withdrawal
by the household, incomplete application), as well as the potential
securitization of the mortgage either by an agency (Fannie Mae, Ginnie
Mae, Freddie Mac, Farmer Mac, thereafter the ``GSEs''), or by private
institutions. The file also reports securitizations independently
of applications. In this paper we do not restrict or filter the file
and include all originations, regardless of their nature. One important
feature of HMDA is the stability of its codebook since 2004, providing
a unique way to assess the evolution of mortgage originations and
securitizations by location over time.

\paragraph*{McDash}

The mortgage data come from the McDash data set compiled by Black
Knight financial using data from the servicing industry. This data
set covers about 75\% of the mortgage market.\footnote{This coverage fluctuates across years, is highest during the housing
boom of 2001-2006, and somewhat lower during the housing bust. Hence
increases in originations in McDash during the latter period are likely
underestimated.} This data source is unique as it provides us with a granular view
of the composition of the mortgage market since 1989, at the 5-digit
ZIP code level. In addition, we obtained exclusive access to 5-digit
ZIP code data, when other researchers have used 3-digit ZIP code data.
An alternative source of data, the HMDA data, is comprehensive within
the reporting rules set by the FFIEC. Yet, such HMDA data does not
include information on mortgage structure (ARM, IO, etc) except in
2018. In McDash we focus on first mortgages in 5-digit ZIP areas with
at least 10 loans, and on mortgages for owner-occupied housing, where
the property value at origination is above \$50,000 and the loan amount
is above \$50,000. Given that the data comes from servicers, we always
use the origination date and not the date where servicing rights were
transferred to prevent double counting. Overall the displayed numbers
are likely underestimates of the true volume of originations in at-risk
areas.

\subsection{Household and Individual Characteristics\label{subsec:Household-and-Individual}}

\paragraph*{American Community Survey}

Household, individual, and housing characteristics are extracted from
the 5-year averages of the American Community Survey, at the Zip code
tabulation area (ZCTA5) level. We use survey weights provided by the
U.S. Census Bureau. The ZCTA5 level was used as the finest level of
geographic disaggregation available in the McDash mortgage data set.
More granular analysis at the census tract level is available.

\subsection{Lender Characteristics}

\paragraph*{The Federal Reserve of Chicago's Commercial Bank data}

Data collected in accordance with the HMDA also provides a mortgage-level
crosswalk with the identity of its lender (RSSDID) in the reporter
panel until 2016 inclusive. Lenders submit a transmittal sheet alongside
the loan application record (LAR). Such transmittal sheet is linked
to the Federal Reserve's RSSDID. The Federal Reserve of Chicago provides
researchers with reports reports of condition and income for all banks
regulated by the Federal Reserve System, Federal Deposit Insurance
Corporation, and the Comptroller of the Currency. As such we do not
observe the liquidity and capital levels of non-bank lenders. We access
such formatted using \possessivecite{drechsler2018banking} consistent
data set.

\section{Six Facts\label{sec:Six-Facts}}

\subsection{Fact \#1: The Volume of Mortgage Originations in Storm Surge Areas
is Rising Sharply}

Table~\ref{tab:hmda_by_flood_risk} computes the total volume of
originations in billions of dollars, in storm surge areas, i.e. more
than 5 feet of storm surge during a Category 4 hurricane, panel (a);
in special flood hazard areas, i.e. in the flood insurance maps 100-year
floodplain, panel (b); in the 3 feet sea level rise area, panel (c);
and in the 6 feet sea level rise area, panel (d).\footnote{The total origination and agency origination volumes are of the same
order of magnitude as those reported by the Urban Institute's Housing
Finance at a Glance series. 2018 is the last year with public disclosure
of HMDA. The 2019 data was not available at the time of writing this
paper.} 

The volume of originations in storm surge areas jumps from 210 billion
dollars in 2012 to 249 billion dollars annually in 2018, i.e. by about
18.5\%. In contrast the total volume of originations declines from
2.1 trillion dollars to 2.0 trillion dollars, a 4.8\% decline. This
decline may be due to changes in reporting requirements. Yet these
changes in reporting requirements also affect the volume of originations
in storm surge areas, whose share in the total volume increases from
9.8\% to 12.5\%, a 2.7 percentage point increase. The volume of agency
originations in storm surge areas is lower in 2018 than in 2012. This
reflects the downward trend in agency originations. Overall, as the
volume of agency originations in storm surge areas declines proportionately
less than the volume of agency originations overall, the share of
agency originations in storm surge areas increases between 2012 and
2018, from 8.6\% to 11\%, a 2.4 percentage point increase.

Similar findings obtain when focusing on areas with more than 10 feet
of storm surge during a category 4 hurricane, instead of areas with
more than 5 feet; or when focusing on category 5 hurricanes. Indeed,
the volume of originations in such areas with more than 10 feet of
surge in a cat. 4 hurricane is 221 billion dollars in 2018 (compared
to 249 billion dollars in areas with more than 5ft of surge). There
is also an increase in originations: from 181 billion dollars of originations
in 2012, to 222 billion dollars of originations in 2018. Similar trends
obtain when focusing on areas with more than 20 feet of storm surge
(from 99 billion dollars in 2012, to 133 billion dollars in 2018).

\subsection{Fact \#2: The Share of Mortgage Originations in Flood Insurance ``Special
Flood Hazard Areas'' is Stable}

This contrasts with the findings obtained when focusing on flood insurance
maps' 100-year floodplains. Table~\ref{tab:hmda_by_flood_risk}(b)
suggests that the volume of originations in Special Flood Hazard Areas
(the 100-year floodplain) is remarkably stable between 2012 and 2018,
at 197.2 billion dollars versus 197 billion dollars, with some fluctuations
in-between these two years. The share of mortgage originations in
SFHAs is also stable, oscillating between 9.2\% and 9.9\% throughout
this period. 

A similar stability obtains when focusing on agency originations in
SFHAs.~The volume of agency originations in SFHAs declines, from
141 billion dollars to 109 billion dollars, with a share in SFHAs
also remarkably stable between 9.2\% and 9.9\%.

\subsection{Fact \#3: Homeowners in Flood Zones Are More Likely to Borrow Using
Non-Traditional Mortgages}

The HMDA mortgage data presented above present a satisfactory aggregate
overview of mortgage originations. They do not provide the structure
of the mortgage, in particular whether the mortgage is a ``simple''
mortgage, fully amortizing with a fixed rate, or whether the mortgage
has any of the complex features described in \citeasnoun{amromin2018complex}:
whether the mortgage is interest-only (and thus does not lead to reductions
in principal amount), whether the mortgage is a fixed or adjustable
rate mortgage (ARM), whether the mortgage was approved using full
document or using low documentations, e.g. for self-employed individuals. 

\begin{figure}
\caption{Evolution of Mortgage Characteristics in the US outside and inside
Storm Surge Areas\label{fig:evolution_complex_mortgages}}

\emph{These figures compare the fraction of complex mortgages for
storm surge areas (at least 5 feet of surge during a Category 4 hurricane)
vs. other areas. }

\begin{center}
\begin{centering}
\subfloat[Share of Interest-Only Mortgages]{\includegraphics[scale=0.4]{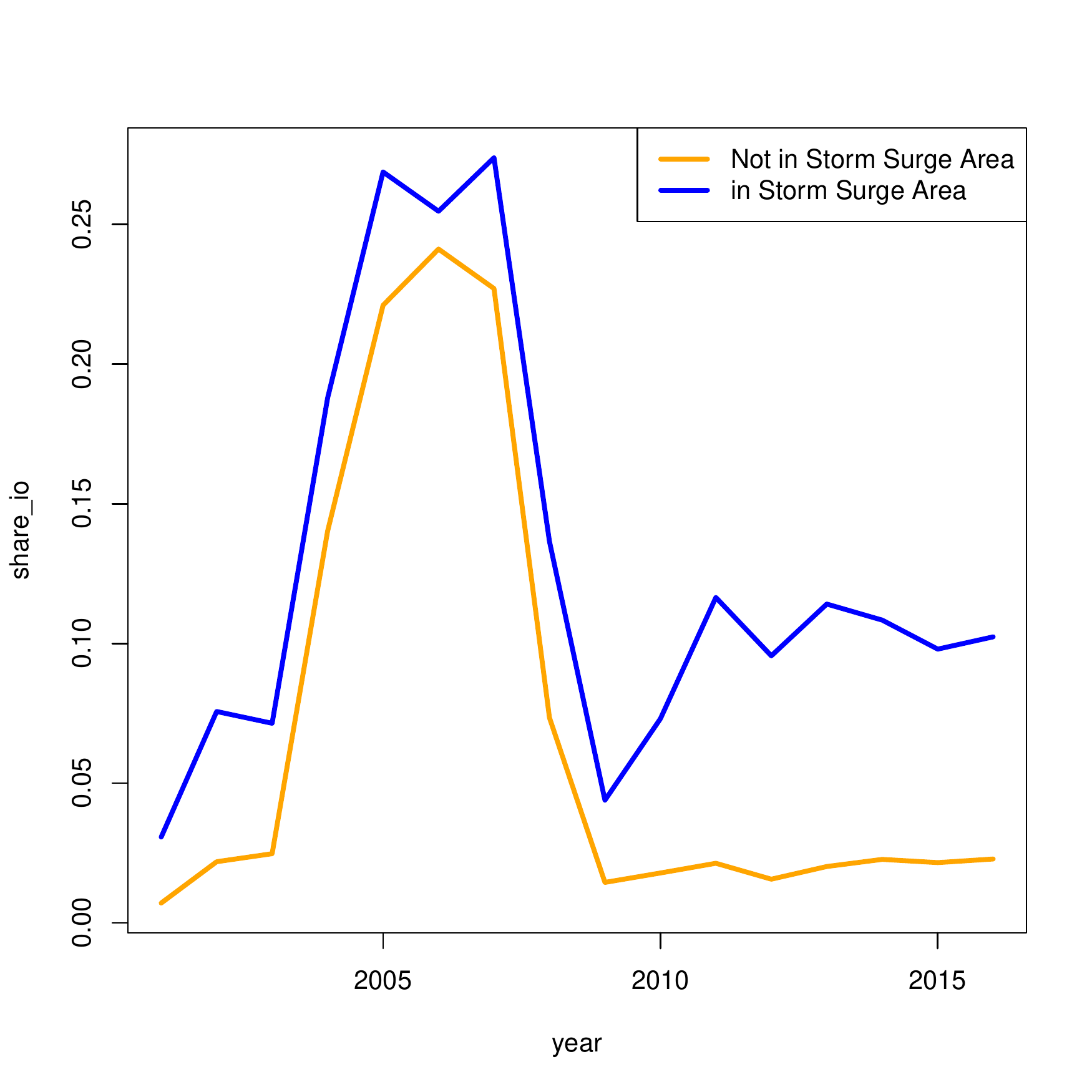} 

}\subfloat[Share of Fixed Rate Mortgages]{\includegraphics[scale=0.4]{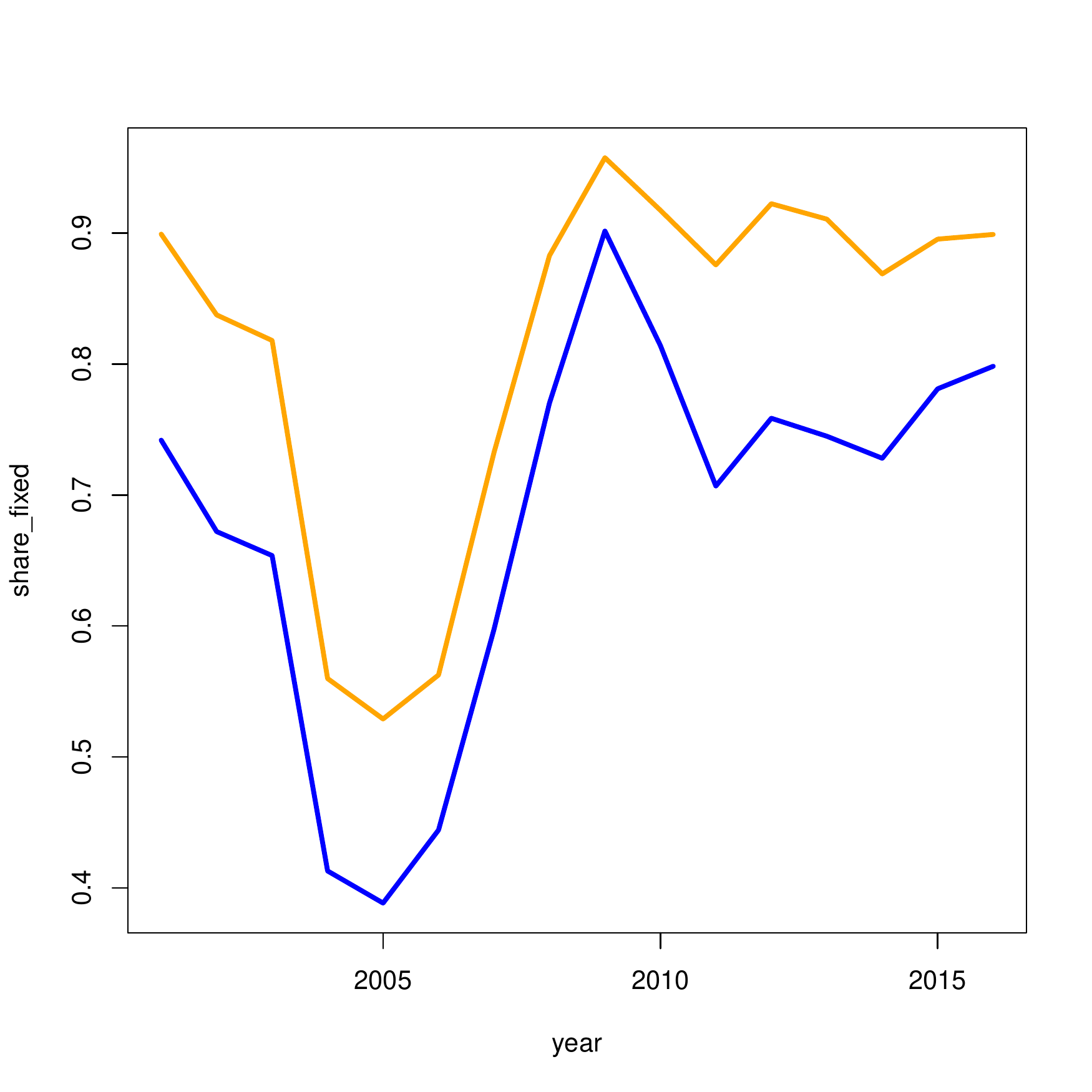} 

}
\par\end{centering}
\begin{centering}
\subfloat[Share Full Documentation]{\includegraphics[scale=0.4]{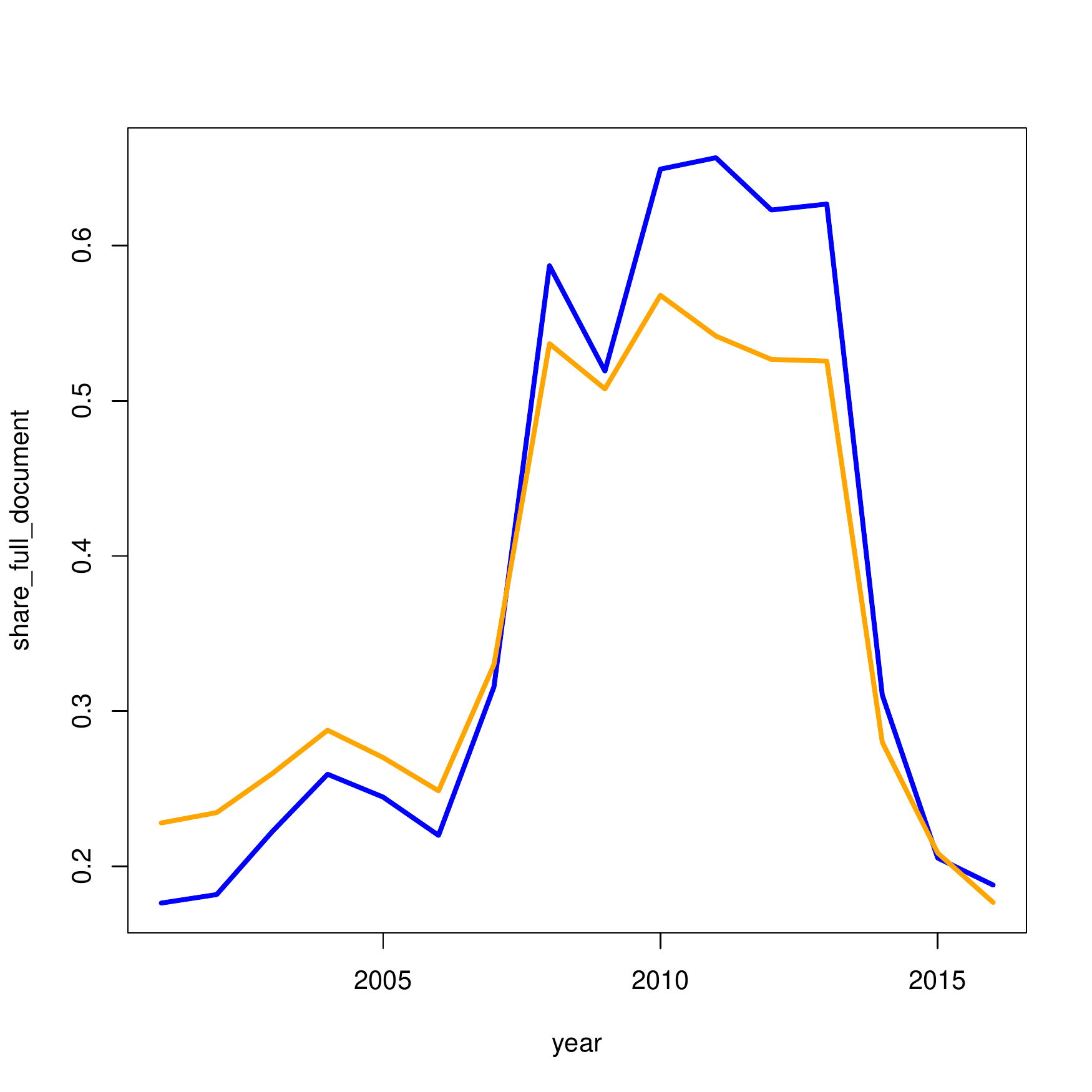} 

}\subfloat[Share No Income No Asset]{\includegraphics[scale=0.4]{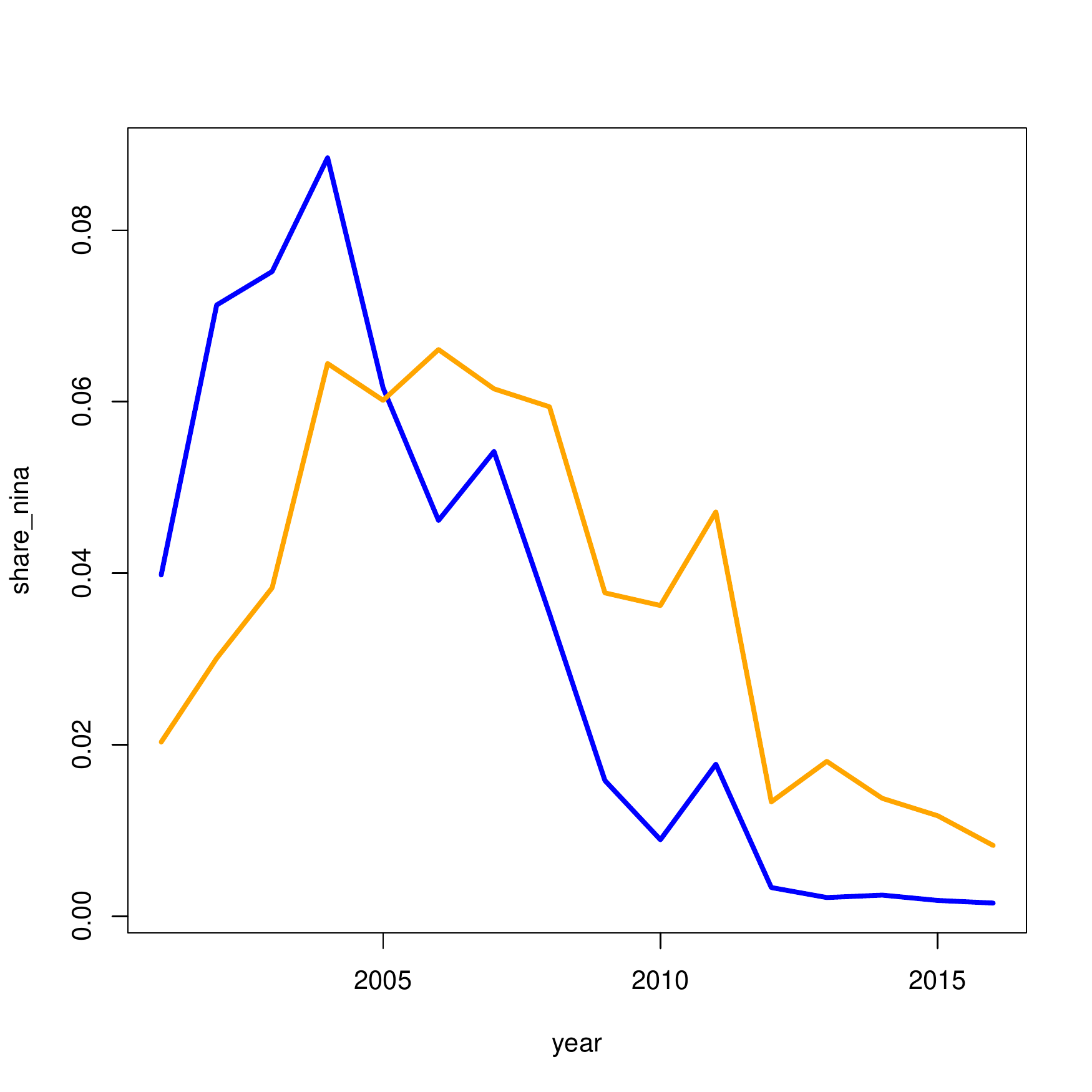} 

}
\par\end{centering}
\end{center}

\emph{Source: Author's calculation using the McDash data set from
Black Knight financial, NOAA's Storm Surge SLOSH simulations, US Census
Bureau shapefiles.}
\end{figure}

Table~\ref{tab:aggregate_statistics} and Figure~\ref{fig:evolution_complex_mortgages}
present aggregate statistics on complex mortgage originations by storm
surge area. As before, a storm surge area is defined here as at least
5 feet of storm surge above ground level during a category 4 hurricane.
A Zip code is in the storm surge area if the maximum surge height
across computation cells is greater than 5 feet. The author has checked
that other definitions of storm surge areas, for instance using 10
feet of surge, or a category 5 hurricane (such as Katrina) yield similar
stylized facts.

The table finds a few such robust stylized facts. First, there is
a persistently higher share of interest-only loans in Storm Surge
Areas: 10.2\% vs 2.3\%. There is a persistently lower share of fixed
rate mortgages in Storm Surge Areas: 79.8\% vs 89.9\%. There are more
full documentation loans and fewer no Income no Asset loans in Storm
Surge Areas vs the rest of the US. This may suggest that while lenders
give more complex mortgages, they ensure that households have better
credit characteristics. The next section suggests that households
in such areas are in fact significantly more vulnerable.

\subsection{Fact \#4: Households in Flood Zones are More Vulnerable}

Table~\ref{tab:ACS2018} compares household, individual, or housing
characteristics in storm surge areas (SLOSH), sea level rise areas,
and Special Flood Hazard areas. Data is from the 5-year average of
the 2018 American Community Survey, at the ZCTA5 level.\footnote{With 36,721 Zip code areas overall, and more than 3,000 per area,
the margins of errors (MOEs) do not affect the significance of the
difference between flood zones in this table.} The ``Rest of the United States'' is made of areas that are in
none of the other three categories. 

Households in Storm Surge areas tend to be poorer than households
in the rest of the U.S., with differences in household income ranging
between \$1,295 (5ft Storm surge vs. rest of the U.S.) and \$2,055
(15ft Storm surge vs. rest of the U.S.). Despite such lower median
incomes, households face higher monthly dollar owner costs, which
translate into higher monthly owner costs as a percentage of income.
Households in storm surge areas tend to be less likely to be living
in owner-occupied housing. Yet we saw in the previous section that
the total volume of mortgage originations is higher than in SFHA areas.
This is due to the higher price of housing, as households in storm
surge areas tend to live in more expensive metropolitan areas.

As the previous analysis suggests, focusing on storm surge areas is
key in describing mortgage vulnerability to coastal flood risk. This
is also the case here for households' vulnerability. Households in
sea level rise areas have significantly higher income, likely related
to the higher amenity value of coastal areas not exposed to the short-term
risk of hurricane storm surges.

The lower panel of the descriptive table (Demographics) shows that
storm surge areas display substantially higher fractions of African
Americans (up to 7.1 ppt higher in storm surge areas vs. the rest
of the United States), higher fractions of Hispanics (18.4\% vs. 17.9\%
in the rest of the United States), and lower fractions of Whites (70.5\%
vs. 76.6\% in the rest of the United States). 

Individuals living in storm surge areas tend to be older, between
0.9 and 2.4 years older than individuals living in the rest of the
United States. While older individuals may have higher savings and
thus may be more resilient, other indicators suggest higher vulnerability.
Individuals in flood risk areas tend to be significantly more likely
to have no health coverage, with up to 15.1\% uninsured in storm surge
areas. The share of individuals below the poverty line is also higher
(+0.4 ppt) in storm surge areas.

Finally, the table also reveals that there is a positive correlation
between vulnerability and the height of a storm surge: when moving
from areas affected by a 5 feet storm surge to an area affected by
a 15 feet storm surge, household income declines (\$63,516 vs \$62,756),
the fraction of individuals below the poverty line increases (14.8\%
vs 15.1\%), the fraction of individuals with no health coverage stays
constant, the fraction of African Americans is broadly stable (20.3\%
vs 20.1\%). Monthly owner costs increase slightly or stay stable as
a percentage of household income (23.4\% vs. 23.5\%). The fraction
of mobile homes increases slightly from 5.53\% to 5.77\%.

\subsection{Fact \#5: Price-to-Rent Ratios Are Declining Relative to the Trend
in Storm Surge Areas, while they are Increasing in Flood Insurance
Areas}

Finally we turn to the evolution of housing markets in storm surge
areas and in FEMA's SFHAs. We proceed by looking at the price-to-rent
ratio as an indicator of future rental risk as the price incorporates
current and future values of rents. In the \citeasnoun{gordon1959dividends}
growth model applied to housing, the price of a real estate asset
reflects its flow of rents net of taxes and maintenance costs, discounted
by the difference of the discount factor and the growth rate of rents.
In this framework, an increase (resp. a decline) in the price to rent
ratio reflects the market's perception of rising (resp. declining)
rents. Rents reflect the current flow utility of amenities. 

If storm surge areas are facing more future risk than flood insurance
SFHAs, we should observe that price-to-rent ratios are experiencing
lower growth than flood insurance SFHAs. If, on the other hand, the
current amenity value of the coast in storm surge areas is similar
to the amenity value of the coast in flood insurance areas, we should
observe no difference in the evolution of rents. 

This is what we test by using Zillow's House Value Index (ZHVI),\footnote{Two other price indices are the Case Shiller index and the FHFA HPI
index (formerly called the OFHEO index). A detailed comparison between
the ZHVI and the Case-Shiller is provided by Zillow in \href{https://wp.zillowstatic.com/3/ZHVI-InfoSheet-04ed2b.pdf}{this document}.} and its Zillow Rental Index (ZRI). Such price index is available
for a subset of 7,439 Zip codes, 339 of which have more than 50\%
of their area in the Special Flood Hazard Area, and 967 of which have
a maximum surge height (at any point of the Zip) above 15 feet. We
consider the period 2012-2018, as in our analysis of mortgage originations.
We then regress the log price to rent, and log rent on a series of
indicator variables for each year; year indicator variables interacted
with a indicator variable for whether the Zip has a maximum surge
height above 15 feet; year indicator variables interacted with an
indicator variable whether the Zip has more than 50\% of its surface
area in an SFHA.

\begin{figure}

\caption{Price-to-Rent Ratio Trends in Storm Surge Areas and in FEMA's Insurance
Map 100-Year Floodplain\label{fig:Price-to-Rent-Ratio-Trends}}

\emph{The upper panel presents the evolution of the price-to-rent
ratio relative to the national trend in Special Flood Hazard Areas
(FEMA's insurance map 100-year floodplain, black points); and the
evolution of the same ratio in storm surge areas (SLOSH model, white
circles). The bars represent 95\% confidence interval, clustering
by Zip code and by year. Vertical axis: 0.05 is 5\%. Red dotted line
for national trend.}

\bigskip{}

\begin{center}

\subfloat[Price-to-Rent Ratio Relative to the National Trend]{

\includegraphics[scale=0.5,trim={0cm 0cm 0cm 2cm},clip]{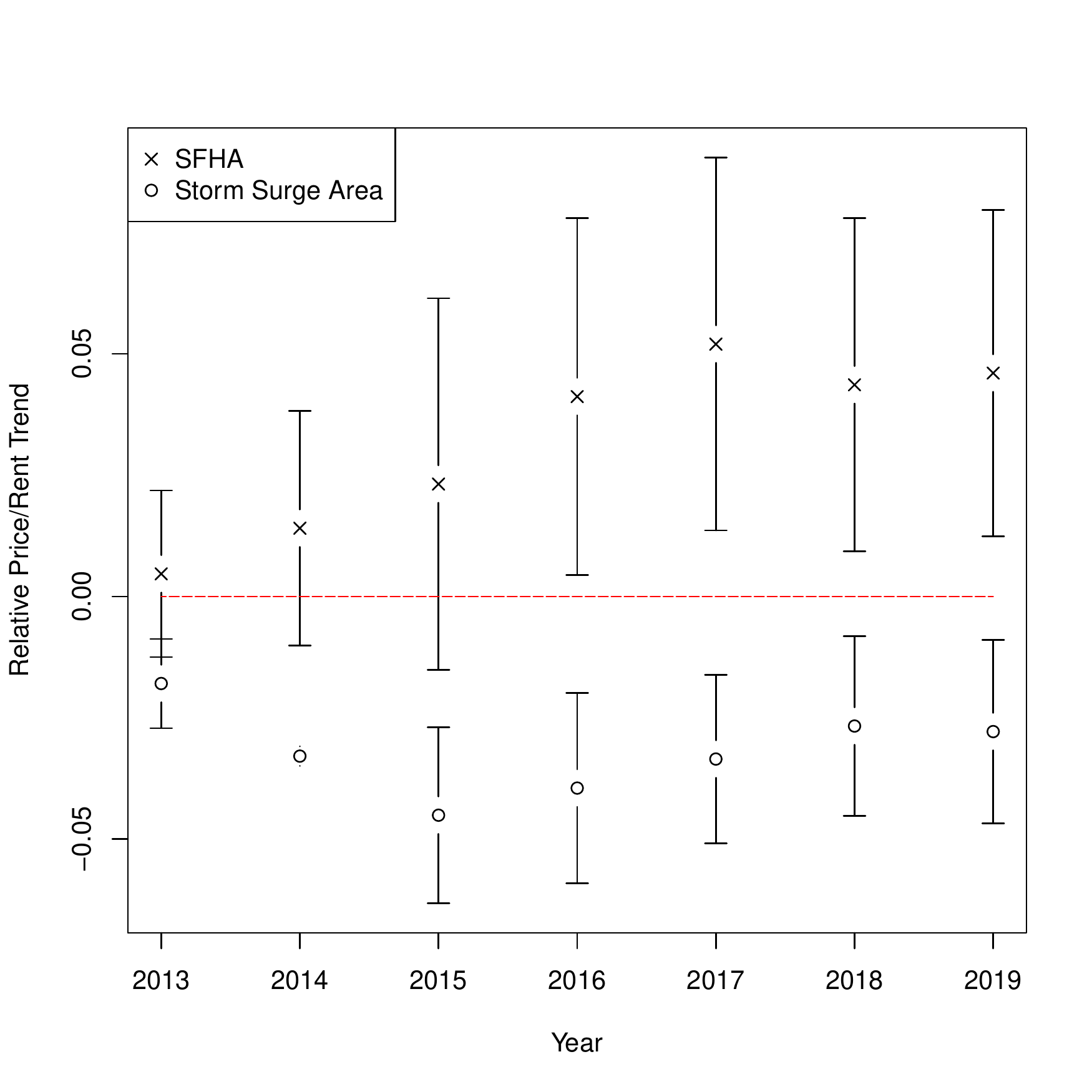}}

\subfloat[Rent Relative to the National Trend]{

\includegraphics[scale=0.5,trim={0cm 0cm 0cm 2cm},clip]{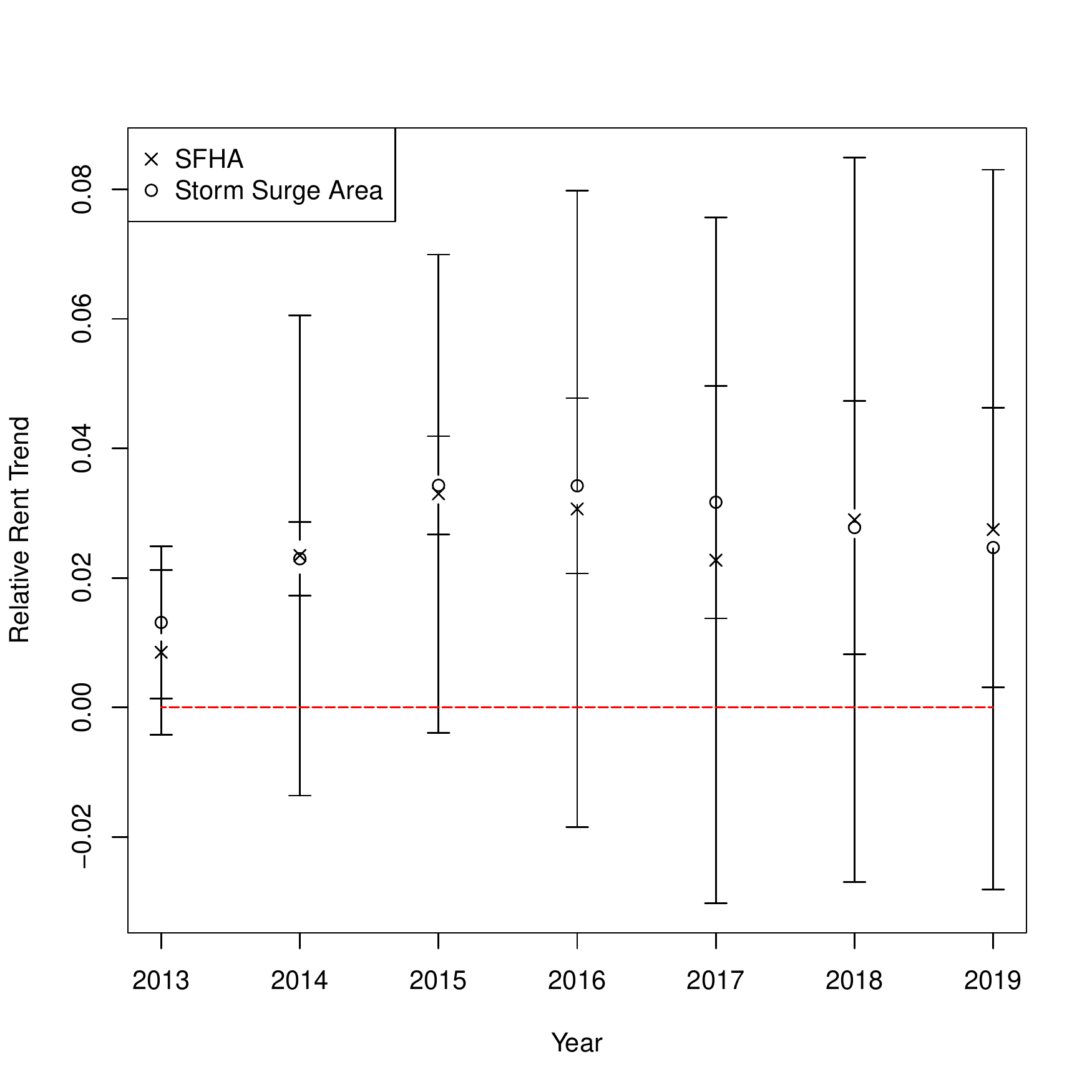}}

\end{center}

\emph{Source: Zillow's ZHVI house price index and ZRI rental index.
Monthly data by postal ZIP.}
\end{figure}

The results presented in the upper panel Figure~\ref{fig:Price-to-Rent-Ratio-Trends}
suggest that price-to-rent ratios are declining significantly (at
95\%) relative to the national house price trends in storm surge areas;
yet they are increasing relative to the national trend in SFHA areas.
The bottom panel suggests that rents evolve similarly at the national
level and in each of these areas. These results suggest that housing
market participants anticipate significantly more risk in storm surge
areas than in flood insurance areas, even as they see no significant
difference in \emph{current} amenity value. 

This supports the hypothesis that storm surge areas may be exposed
to risk that is not typically measured when focusing on flood insurance
areas.

\subsection{Fact \#6: Lenders in Storm Surge Flood Zones Are More Likely to Be
Large Commercial Banks}

Table~\ref{tab:Lenders-in-Storm} presents the characteristics and
key ratios of lenders in storm surge areas and in flood insurance
SFHA areas. The ratios are computed using the first quarter of 2012,
the beginning of the expansion of mortgage credit in storm surge areas
in Table~\ref{tab:hmda_by_flood_risk}. Each ratio is obtained by
taking the mean of bank characteristics weighted by their volume of
mortgage originations in the area. These ratios and characteristics
are available for bank lenders. The last row of the table presents
the share of non-bank lenders in the sample. The non-bank origination
share is large overall, consistent with the evidence of \citeasnoun{goodman2020housing}. 

The evidence presented in this table supports the hypothesis that
lenders in storm surge areas tend to be larger banks (about 3\% larger,
719b\$ vs. 672\$) than in flood insurance areas. This is true both
on average and for the median bank lender. There is a positive relationship
between the storm surge height and the size of the bank lender. Such
monotonicity between bank lender characteristics and surge height
is typical of the findings of this paper. 

Consistent with this larger size, the table suggests that banks in
storm surge areas have marginally lower return on assets (0.238\%
vs. 0.243\%) and marginally lower return on equity (2.26\% vs. 2.3\%).
They tend to rely less on loans in their portfolio~(58.2\% in loans
over assets vs. 58.4\%), and they have similar leverage (equity over
assets of 10.8\% in all four zones). 

These findings are substantially different than the ones from \citeasnoun{beltratti2012credit}.
In their work, they found that banks that performed better between
July 2007 and December 2008 tended to have lower returns and lower
leverage. This paper finds that banks with a portfolio of mortgage
loans most exposed to hurricane storm surges tend to have lower returns
and similar leverage. One common finding with \citeasnoun{beltratti2012credit}
is that banks exposed to storm surges tend to rely less on their loan
portfolio.

\section{Conclusion\label{sec:Conclusion}}

This paper presents an assessment of coastal flood risk in the mortgage
market that relies on scientific simulations of storm surges rather
than on flood insurance maps, which are inherently dependent on a
community's decision to participate in the program. Relying on independent
measures of storm surge risk presents a substantially different picture
of coastal flood risk: there is a large and rising volume of mortgages
at risk (constant in flood insurance areas), households are more likely
to be African American or Hispanic, house prices exhibit lower trends,
and lenders tend to be larger and are more likely to be commercial
banks.

These findings suggest the importance of relying on measures of flood
risk that are independent from flood insurance maps. Future research
could explore the endogeneity of flood insurance mapping to the demographics
of the neighborhoods, conditional on objective flood risk measures.
Future research could also explore the optimal risk taking of financial
institutions facing ambiguous flood risk in coastal areas. 

\bibliographystyle{agsm}
\bibliography{aggregate_climate_risk_in_the_mortgage_market}

\clearpage{}\pagebreak{}

\clearpage{}\pagebreak{}

\clearpage{}\pagebreak{}

\clearpage{}\pagebreak{}

\clearpage{}\pagebreak{}

\begin{table}

\caption{Aggregate Statistics on Mortgage Originations by Flood Risk Area \textendash{}
Home Mortgage Disclosure Act Data\label{tab:hmda_by_flood_risk}}

\emph{These four panels present the unweighted sum of mortgage originations
in the HMDA data, for four different types of flood zones: storm surge
(at least 5 feet in a cat. 4 hurricane), flood insurance areas (SFHAs),
and 3-6ft sea level rise. Data includes all originations. Agency originations:
from Fannie Mae, Freddie Mac, Ginnie Mae, and Farmer Mac. }

\subfloat[Storm Surge Areas]{

{\footnotesize
\begin{tabular}{P{2cm}P{2cm}P{2cm}P{2cm}P{2cm}P{2cm}P{2cm}}
  \hline
Year & Origination Amount (b\$) & In Surge Area (b\$) & In Surge Area (\%) & Agency Amount (b\$) & Agency in Surge Area (b\$) & Agency in Surge Area (\%) \\ 
  \hline
2012 & 2,135 & 210 & 9.80 & 1,528 & 131 & 8.60 \\ 
  2013 & 1,903 & 203 & 10.60 & 1,609 & 159 & 9.90 \\ 
  2014 & 1,386 & 151 & 10.90 &   826 &  77 & 9.30 \\ 
  2015 & 1,848 & 197 & 10.70 & 1,138 & 105 & 9.20 \\ 
  2016 & 2,181 & 225 & 10.30 & 1,375 & 125 & 9.10 \\ 
  2017 & 1,930 & 211 & 10.90 & 1,170 & 115 & 9.80 \\ 
  2018 & 1,993 & 249 & 12.50 & 1,100 & 121 & 11.00 \\ 
   \hline
\end{tabular}

}}

\subfloat[Special Flood Hazard Areas]{

{\footnotesize
\begin{tabular}{P{2cm}P{2cm}P{2cm}P{2cm}P{2cm}P{2cm}P{2cm}}
  \hline
Year & Origination Amount (b\$) & In SFHA (b\$) & In SFHA (\%) & Agency Amount (b\$) & Agency in SFHA (b\$) & Agency in SFHA (\%) \\ 
  \hline
2012 & 2,135 & 197 & 9.20 & 1,528 & 141 & 9.24 \\ 
  2013 & 1,903 & 180 & 9.50 & 1,609 & 152 & 9.48 \\ 
  2014 & 1,386 & 134 & 9.70 &   826 &  79 & 9.66 \\ 
  2015 & 1,848 & 177 & 9.60 & 1,138 & 108 & 9.57 \\ 
  2016 & 2,181 & 207 & 9.50 & 1,375 & 130 & 9.51 \\ 
  2017 & 1,930 & 190 & 9.90 & 1,170 & 115 & 9.85 \\ 
  2018 & 1,993 & 198 & 9.90 & 1,100 & 109 & 9.93 \\ 
   \hline
\end{tabular}

}}

\subfloat[3ft Sea Level Rise Areas]{

{\footnotesize
\begin{tabular}{P{2cm}P{2cm}P{2cm}P{2cm}P{2cm}P{2cm}P{2cm}}
  \hline
Year & Origination Amount (b\$) & In 3 ft SLR (b\$) & In 3 ft SLR (\%) & Agency Amount (b\$) & Agency in 3ft SLR (b\$) & Agency in 3ft SLR (\%) \\ 
  \hline
  2012 & 2,135 & 34 & 1.60 & 1,528 & 21 & 1.37 \\ 
  2013 & 1,903 & 31 & 1.60 & 1,609 & 23 & 1.43 \\ 
  2014 & 1,386 & 23 & 1.60 &   826 & 11 & 1.33 \\ 
  2015 & 1,848 & 30 & 1.60 & 1,138 & 15 & 1.32 \\ 
  2016 & 2,181 & 35 & 1.60 & 1,375 & 18 & 1.31 \\ 
  2017 & 1,930 & 32 & 1.60 & 1,170 & 16 & 1.36 \\ 
  2018 & 1,993 & 33 & 1.70 & 1,100 & 15 & 1.37 \\ 
   \hline
\end{tabular}

}
}

\subfloat[6ft Sea Level Rise Areas]{

{\footnotesize
\begin{tabular}{P{2cm}P{2cm}P{2cm}P{2cm}P{2cm}P{2cm}P{2cm}}
  \hline
Year & Origination Amount (b\$) & In 6 ft SLR (b\$) & In 6 ft SLR (\%) & Agency Amount (b\$) & Agency in 6ft SLR (b\$) & Agency in 6ft SLR (\%) \\ 
  \hline
2012 & 2,135 & 67 & 3.10 & 1,528 & 42 & 2.75 \\ 
2013 & 1,903 & 64 & 3.30 & 1,609 & 46 & 2.86 \\ 
  2014 & 1,386 & 46 & 3.30 &   826 & 22 & 2.66 \\ 
  2015 & 1,848 & 62 & 3.30 & 1,138 & 31 & 2.72 \\ 
  2016 & 2,181 & 71 & 3.30 & 1,375 & 37 & 2.69 \\ 
  2017 & 1,930 & 64 & 3.30 & 1,170 & 32 & 2.74 \\ 
  2018 & 1,993 & 70 & 3.50 & 1,100 & 31 & 2.81 \\ 
   \hline
\end{tabular}

} 
}

\emph{Source: author's calculations from FFIEC's HMDA, NOAA's SLOSH,
FEMA's NFHL, NOAA's Sea Level Rise layers, and US Census Bureau shapefiles. }
\end{table}
\begin{table}
\caption{Aggregate Statistics on Mortgage Originations, by Mortgage Type \textendash{}
McDash data set\label{tab:aggregate_statistics}}

\emph{Storm surge area: at least 5 feet of storm surge above ground
level during a Category 4 hurricane at high tide. Sample: First mortgages,
in 5-digit ZIP areas with at least 10 loans, and on mortgages for
owner-occupied housing, where the property value at origination is
above \$50,000 and the loan amount is above \$50,000. }

\bigskip{}

\begin{center}

\subfloat[Storm Surge Areas]{{\footnotesize 
\begin{tabular}{P{1.2cm}P{1.2cm}P{1.2cm}P{1.2cm}P{1.2cm}P{1.2cm}P{1.2cm}P{1.2cm}P{1.2cm}P{1.2cm}}
  \hline
  Year & Total & Interest Only  & Interest Only  & Fixed Rate  & Fixed Rate& Full Doc.  & Full Doc.  & N.I.N.A.  & N.I.N.A. \\
       & M\$         & M\$ & \% & M\$ & \% & M\$ & \% & M\$ & \% \\
  \hline
2006 & 130,058 & 33,128 & 25.5\% & 57,775 & 44.4\% & 28,617 & 22.0\% & 6,003 & 4.6\% \\ 
  2007 &  89,887 & 24,617 & 27.4\% & 53,698 & 59.7\% & 28,386 & 31.6\% & 4,871 & 5.4\% \\ 
  2008 &  46,444 &  6,341 & 13.7\% & 35,770 & 77.0\% & 27,268 & 58.7\% & 1,638 & 3.5\% \\ 
  2009 &  49,223 &  2,162 &  4.4\% & 44,377 & 90.2\% & 25,555 & 51.9\% &   779 & 1.6\% \\ 
  2010 &  41,824 &  3,056 &  7.3\% & 34,050 & 81.4\% & 27,157 & 64.9\% &   373 & 0.9\% \\ 
  2011 &  37,329 &  4,348 & 11.6\% & 26,386 & 70.7\% & 24,512 & 65.7\% &   662 & 1.8\% \\ 
  2012 &  54,484 &  5,208 &  9.6\% & 41,330 & 75.9\% & 33,942 & 62.3\% &   182 & 0.3\% \\ 
  2013 &  51,983 &  5,933 & 11.4\% & 38,721 & 74.5\% & 32,583 & 62.7\% &   113 & 0.2\% \\ 
  2014 &  34,266 &  3,713 & 10.8\% & 24,947 & 72.8\% & 10,639 & 31.0\% &    84 & 0.2\% \\ 
  2015 &  51,535 &  5,050 &  9.8\% & 40,253 & 78.1\% & 10,587 & 20.5\% &    95 & 0.2\% \\ 
  2016 &  56,364 &  5,771 & 10.2\% & 44,994 & 79.8\% & 10,596 & 18.8\% &    87 & 0.2\% \\ 
   \hline
\end{tabular}

}   }

\bigskip{}
\subfloat[Other ZIP Codes]{

{\footnotesize
\begin{tabular}{P{1.2cm}P{1.2cm}P{1.2cm}P{1.2cm}P{1.2cm}P{1.2cm}P{1.2cm}P{1.2cm}P{1.2cm}P{1.2cm}}
  \hline
    Year & Total & Interest Only  & Interest Only  & Fixed Rate  & Fixed Rate& Full Doc.  & Full Doc.  & N.I.N.A.  & N.I.N.A. \\
       & M\$         & M\$ & \% & M\$ & \% & M\$ & \% & M\$ & \% \\
  \hline
2006 & 1,685,407 & 406,480 & 24.1\% &   948,216 & 56.3\% & 419,192 & 24.9\% & 111,374 & 6.6\% \\ 
  2007 & 1,429,807 & 324,624 & 22.7\% & 1,047,208 & 73.2\% & 472,041 & 33.0\% &  87,932 & 6.1\% \\ 
  2008 &   903,184 &  66,331 &  7.3\% &   797,292 & 88.3\% & 484,876 & 53.7\% &  53,654 & 5.9\% \\ 
  2009 & 1,148,235 &  16,648 &  1.4\% & 1,099,416 & 95.7\% & 582,970 & 50.8\% &  43,295 & 3.8\% \\ 
  2010 &   984,994 &  17,594 &  1.8\% &   903,601 & 91.7\% & 559,498 & 56.8\% &  35,691 & 3.6\% \\ 
  2011 &   851,573 &  18,178 &  2.1\% &   745,714 & 87.6\% & 461,347 & 54.2\% &  40,153 & 4.7\% \\ 
  2012 & 1,293,600 &  20,246 &  1.6\% & 1,193,174 & 92.2\% & 681,359 & 52.7\% &  17,253 & 1.3\% \\ 
  2013 & 1,043,271 &  21,058 &  2.0\% &   950,015 & 91.1\% & 548,348 & 52.6\% &  18,840 & 1.8\% \\ 
  2014 &   594,731 &  13,523 &  2.3\% &   516,695 & 86.9\% & 166,558 & 28.0\% &   8,177 & 1.4\% \\ 
  2015 &   793,719 &  17,128 &  2.2\% &   710,655 & 89.5\% & 165,605 & 20.9\% &   9,299 & 1.2\% \\ 
  2016 &   811,134 &  18,553 &  2.3\% &   729,140 & 89.9\% & 143,375 & 17.7\% &   6,694 & 0.8\% \\ 
   \hline
\end{tabular}

} }

\end{center}

\bigskip{}

\emph{Source: Author's calculations using the McDash data provided
by Black Knight financial.}
\end{table}

\clearpage{}\pagebreak{}

\begin{table}
\caption{Census Demographics in Storm Surge Areas\label{tab:ACS2018}}

\emph{This table presents summary statistics for households and individuals
in Zip code tabulation areas (ZCTAs), depending on storm surge heights
in the ZIP code tabulation area (columns 2\textendash 4) and depending
on the share of the ZIP code in Special Flood Hazard Areas, the 100-year
floodplain of insurance maps.}

\bigskip
\begin{center} 
{\footnotesize  
\captionsetup[table]{labelformat=empty,skip=1pt}
\begin{longtable}{p{4.1cm}P{2cm}P{2cm}P{2cm}P{2cm}P{2cm}}
\toprule
& (1) & (2) & (3) & (4) & (5) \\
& Rest of the U.S. & Storm Surge 5ft, Cat 4 Hurricane & Storm Surge 10ft, Cat 4 Hurricane & Storm Surge 15ft, Cat 4 Hurricane & Special Flood Hazard Area \\ 
\midrule
\\
\multicolumn{4}{l}{\emph{Housing Units}} \\
\\
House value $\ddagger$ & \$252,288 & \$317,241 & \$315,413 & \$318,604 & \$228,930 \\ 
Monthly owner cost $\ddagger$ $\star$ & \$1,226.1 & \$1,360.8 & \$1,347.1 & \$1,347 & \$1,160.3 \\ 
As \% of income $\ddagger$ $\star$ & 21.8\% & 23.4\% & 23.4\% & 23.5\% & 21.8\% \\ 
Gross rent $\ddagger$ & \$1,055.5 & \$1,244.6 & \$1,241.8 & \$1,232.5 & \$1,047.1 \\ 
\% Owner occupied & 64.8\% & 59.5\% & 59.5\% & 59.0\% & 65.1\% \\
\% Mobile homes & 5.93\% & 5.53\% & 5.7\% & 5.77\% & 7.19\% \\
\\
\multicolumn{4}{l}{\emph{Demographics}} \\
\\
Household income $\ddagger$ & \$64,811 & \$63,516 & \$63,205 & \$62,756 & \$62,698 \\ 
Age $\ddagger$ & 38.5 & 39.8 & 39.9 & 39.9 & 39.3 \\ 
\% Asian & 6.21\% & 5.57\% & 5.5\% & 5.45\% & 4.86\% \\ 
\% Black & 13.2\% & 20.3\% & 20.1\% & 20.1\% & 15.6\% \\ 
\% White & 76.6\% & 70.5\% & 70.5\% & 70.1\% & 76.7\% \\ 
\% Hispanic & 17.9\% & 18.4\% & 18.2\% & 18.4\% & 16.9\% \\ 
\% Below poverty & 14.4\% & 14.8\% & 14.9\% & 15.1\% & 14.6\% \\ 
\% No health coverage & 12.9\% & 15.1\% & 15.0\% & 15.1\% & 14.4\% \\
\\
\midrule
\\
Zip code areas & 36721 & 36721 & 36721 & 36721 & 36721 \\ 
Share of population in area & 89.8\% & 10.2\% & 9.04\% & 7.21\% & 12.1\% \\
\\
\bottomrule
\end{longtable}

}   
\end{center}  
\bigskip  

\emph{$\ddagger$: median. $\star$: households with a mortgage. Source:
ZCTA5 5-year average of the 2018 American Community Survey. NOAA's
Sea LevelrRise layer, NOAA's SLOSH MOMs, FEMA's Special Flood Hazard
Areas from the 2017 National Flood Hazard Layer (NFHL).}
\end{table}

\clearpage{}\pagebreak{}
\begin{table}
\caption{Lenders in Storm Surge Areas\label{tab:Lenders-in-Storm}}

\emph{This table presents bank lenders' key ratios in four types of
flood zones (defined before). Each ratio is the average ratio of bank
lenders in each zone in the first quarter of 2012, weighted by lenders'
mortgage origination volume in that zone.}

\bigskip{}

\begin{center}
\begin{tabular}{lcccc}
  \toprule
  & \multicolumn{4}{c}{Flood Zone:} \\
 & (1) & (2) & (3) & (4) \\ 
  \cmidrule(lr){2-5}
  & Storm Surge 5ft & Storm Surge 10ft & Storm Surge 20ft & SFHA \\
  \cmidrule(lr){2-5}
  \\
  Average Assets (M\$) & 719,933 & 717,457 & 743,404 & 672,091 \\
  \\
  Median Assets (M\$) & 330,227 & 330,227 & 330,227 & 206,808 \\
  \\
  Return on Assets (Quarterly) & 0.238\% & 0.238\% & 0.233\% & 0.243\% \\
  \\
  Return on Equity (Quarterly) & 2.26\% & 2.26\% & 2.22\% & 2.3\% \\
  \\
  Loans over Assets & 58.2\% & 58.2\% & 57.5\% & 58.4\% \\
  \\
  Deposits over Assets & 69.7\% & 69.7\% & 69.1\% & 71.2\% \\
  \\
  Liquidity over Assets & 19.3\% & 19.2\% & 19.3\% & 19.9\% \\
  \\
  Equity over Assets & 10.8\% & 10.8\% & 10.8\% & 10.8\% \\ 
  \\
  Share of Non-Bank Lenders & 62.66\% & 62.35\% & 60.0\% & 62.9\% \\
  \\
 \bottomrule
\end{tabular}
 
\end{center}

\bigskip{}

\emph{Source: Calculations of the author from commercial bank data
from the Federal Reserve of Chicago and cleaned by \citeasnoun{drechsler2018banking}. }
\end{table}

\end{document}